\documentclass[prl,superscriptaddress,floatfix,nofootinbib,twocolumn,aps,10pt]{revtex4-2}
\usepackage{graphicx,amssymb,amsmath,bm,dcolumn,xcolor,mathtools}
\usepackage[caption=false]{subfig}
\usepackage[colorlinks,linkcolor=black,citecolor=blue,urlcolor=blue]{hyperref}
\usepackage{url}
\usepackage{upgreek}
\usepackage{threeparttable}
\usepackage{comment}
\usepackage{xfrac}
\usepackage{appendix}
\usepackage{enumitem}
\usepackage[normalem]{ulem}
\usepackage{soul}
\usepackage{multirow}
\newcounter{parnum}
\begin{document}

\title{Microscale torsion resonators for short-range gravity experiments}

\author{J. Manley}
\affiliation{Wyant College of Optical Sciences, University of Arizona, Tucson, AZ 85721, USA}

\author{C. A. Condos}
\affiliation{Wyant College of Optical Sciences, University of Arizona, Tucson, AZ 85721, USA}

\author{S. Schlamminger}
\affiliation{National Institute of Standards and Technology, Gaithersburg, MD 20899, USA}

\author{J. R. Pratt}
\affiliation{National Institute of Standards and Technology, Gaithersburg, MD 20899, USA}

\author{D. J. Wilson}
\affiliation{Wyant College of Optical Sciences, University of Arizona, Tucson, AZ 85721, USA}

\author{W. A. Terrano}
\affiliation{Department of Physics, Arizona State University, Tempe, AZ 85281, USA}

\begin{abstract}
Measuring gravitational interactions on sub-100-$\upmu$m length scales offers a window into physics beyond the Standard Model. However, short-range gravity experiments are limited by the ability to position sufficiently massive objects to within small separation distances. Here we propose mass-loaded silicon nitride ribbons as a platform for testing the gravitational inverse square law at separations currently inaccessible with traditional torsion balances. These microscale torsion resonators benefit from low thermal noise due to strain-induced dissipation dilution while maintaining compact size ($<100~\upmu$g) to allow close approach. Considering an experiment combining a $40~\upmu$g torsion resonator with a source mass of comparable size ($130~\upmu$g) at separations down to 25$~\upmu$m, and including limits from thermomechanical noise and systematic uncertainty, we predict these devices can set novel constraints on Yukawa interactions within the 1-100$~\upmu$m range.
\end{abstract}

\maketitle

Measurements of the gravitational inverse square law (ISL) at short distances can test physics beyond the Standard Model~\cite{adelberger2003tests,adelberger2009torsion}. Dark energy generically suggests a scale for new physics around 100$~\upmu$m~\cite{Kapner2007}, and several specific theories have been proposed in this regime. For example, ``fat gravitons" with size in the 20-95$~\upmu$m range may solve the cosmological constant problem~\cite{sundrum2004fat,adelberger2007implications}, chameleon fields can have interaction lengths as low as  ${\sim25}~\upmu$m in a laboratory setting~\cite{upadhye2012chameleon}, and modifications to gravity from a ``dark dimension” are predicted to arise between 1-10$~\upmu$m~\cite{montero2023dark,vafa2024swamplandish}. Experimental tests of the ISL are also sensitive to short-range Yukawa interactions mediated by massive particles such as dilatons~\cite{kaplan2000couplings}, radions~\cite{antoniadis2003brane}, or gauge bosons~\cite{arkani1999phenomenology,dimopoulos2003probing}. In this context, one can parametrize a violation of the ISL with a Yukawa potential that augments the Newtonian gravitational potential as
\begin{equation} \label{eq:yukawapotential}
    V(r) = - \frac{G M}{r} \left(1 + \alpha {e^{-r/\lambda}}\right)
\end{equation}
at a distance $r$ from a point-mass source $M$. Here $\alpha$ is the interaction strength relative to Newtonian gravity and $\lambda=\hbar/(m_\text{b} c)$ is the interaction length, where $m_\text{b}$ is the mass of the exchange boson.

A variety of devices have been used to search for Yukawa interactions in the sub-100-$\upmu$m range, including torsion balances~\cite{Hoyle2001,hoyle2004submillimeter, Kapner2007, lee2020new}, optomechanical cantilevers~\cite{geraci2008improved}, microelectromechanical torsion oscillators~\cite{chen2016stronger}, levitated particles~\cite{blakemore2021search}, and Casimir experiments~\cite{fischbach2001new,mostepanenko2001constraints}. Of these, only cm-scale torsion balances with test masses exceeding 100~mg have achieved sensitivity to gravitational-strength ($\left|\alpha\right|=1$) interactions~\cite{Kapner2007,tu2007null,Cook2013,tan2016new,lee2020new,tan2020improvement}. The upper bound for the interaction length of gravitational-strength Yukawa interactions has been progressively constrained by the E\"{o}t-Wash group from $197~\upmu$m in 2001~\cite{Hoyle2001,hoyle2004submillimeter} to $56~\upmu$m in 2007~\cite{Kapner2007}, to 42$~\upmu$m in 2013~\cite{Cook2013}, and most recently to 39$~\upmu$m in 2020~\cite{lee2020new}. Device planarity and the onset of electrostatic noise set the lower limit on the minimum surface separation between the source and test masses in torsion balance experiments~\cite{lee2020new,tan2020improvement}, making it difficult to probe shorter interaction lengths.

\begin{figure}[b]
    \centering
    \includegraphics[width=\columnwidth,trim= 0in 0in 0in 0in]{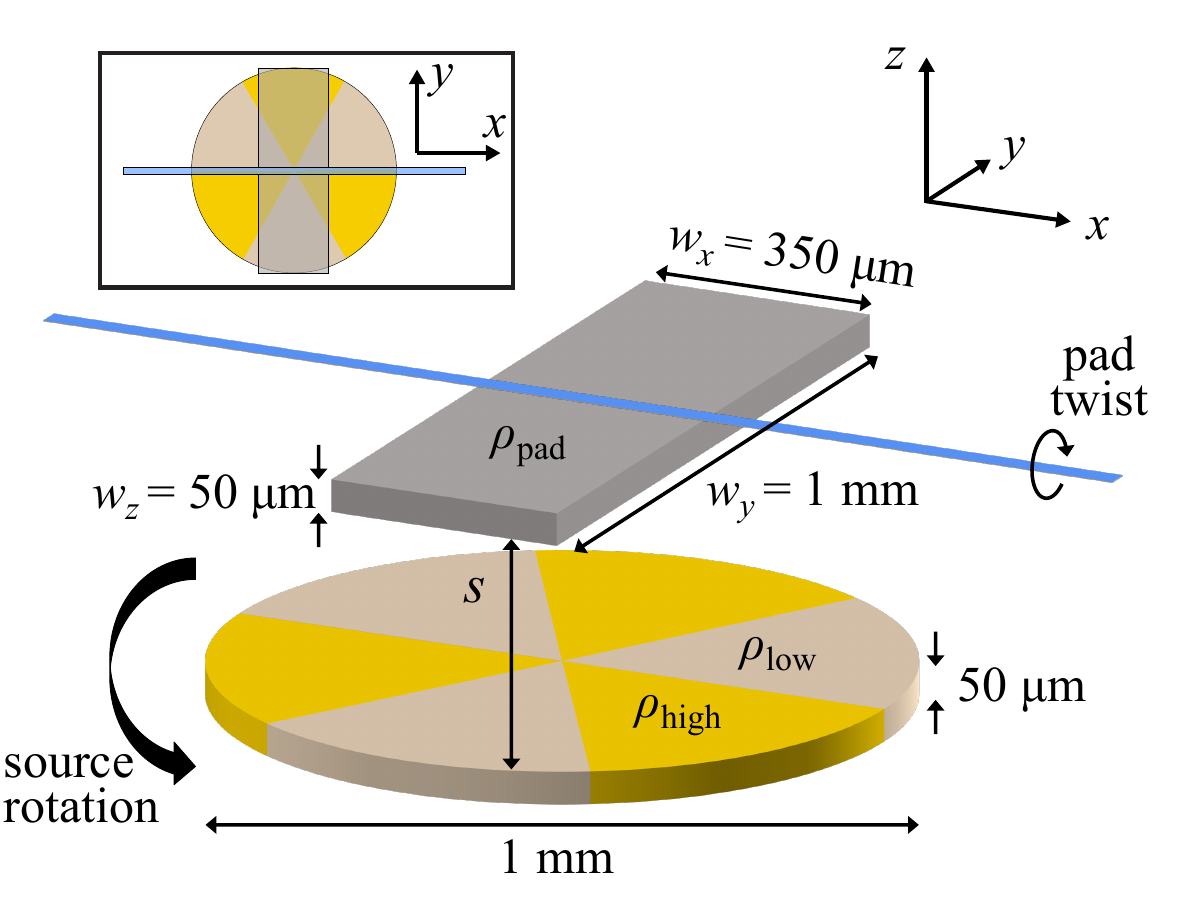}
    \caption{\textbf{Sketch of the proposed experiment.} A $40~\upmu$g silicon pad is suspended from a Si$_3$N$_4$ nanoribbon, forming a torsion resonator. The heterogeneous source mass consists of an alternating density pattern, exerting a gravitational torque (along the $x$-axis) on the pad. The high-density $\rho_\text{high}$ region could for example consist of micromachined tungsten or gold ($\rho_\text{Au}\approx \rho_\text{W}=1.9\times 10^4 ~\text{kg}/{\rm m}^3$) that is electroplated on a silicon wafer, such that each wedge would be $130~\upmu$g. While the low-density $\rho_\text{low}$ region could be empty space, it is common to use another material of minimal density such as silicon~\cite{geraci2008improved} or epoxy~\cite{weld2008new} to maintain a planar surface.  An electrostatic shield (not shown) separates the pad and attractor. Inset: Top view showing the attractor orientation that produces maximal torque. }
	\label{fig:diagram}
\end{figure}

\begin{figure*}[t]
    \centering
    \includegraphics[width=2\columnwidth]{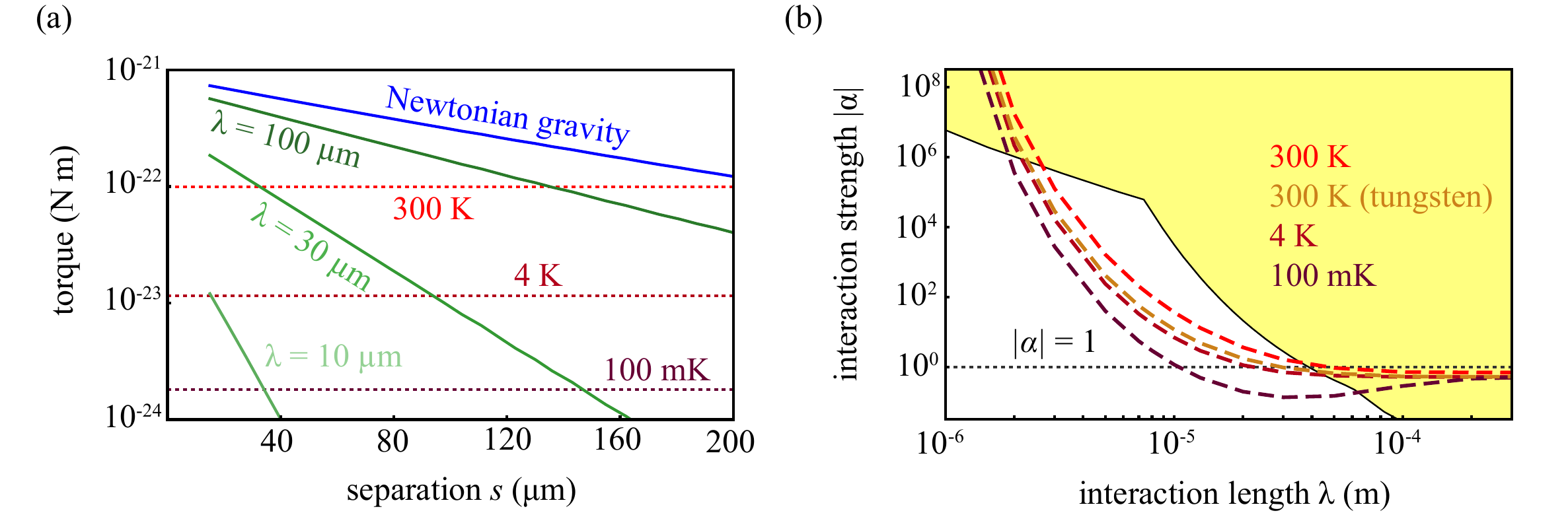}
    \caption{ \textbf{Projections for an ISL experiment.}  (a) Expected Newtonian and Yukawa (for $\alpha=1$) torques, calculated numerically for the orientation depicted in the inset of Fig. \ref{fig:diagram} assuming $\rho_\text{low}=0$ and $\rho_\text{high}=\rho_\text{Au}$. The dashed lines indicate the expected thermal noise-equivalent torque $\delta {N}$ for a single measurement over an averaging time $t_\text{meas}=1$~day at various device temperatures. (b) Dashed curves indicate the projected 2$\sigma$ minimum detectable coupling strength $\alpha_\text{min}$, assuming a measurement campaign that consists of 30 torque measurements, lasting a day each, made at various separations in the range $[25,75]~ \upmu$m. Each projection accounts for limitations from thermal noise and systematic uncertainties. The yellow shaded region corresponds to existing constraints from Refs. \cite{chen2016stronger,lee2020new,tan2020improvement}. All calculations assume the pad to be composed of silicon, $\rho_\text{pad}=2330 ~\text{kg}/{\rm m}^3$, except for an additional projection in (b) accounting for a case where the pad is fabricated from a denser material (tungsten), $\rho_\text{pad} = \rho_\text{W}$.}
	\label{fig:ISL_projections}
\end{figure*}

Here we propose microscale torsion resonators as a platform for measuring gravity on sub-100-$\upmu$m length scales. Introduced in Ref.~\cite{pratt2023nanoscale}, these devices are based on strained Si$_3$N$_4$ nanoribbons that have been mass-loaded with a silicon pad to form sub-milligram, chip-scale torsion resonators. The compactness and optically flat planar geometry of the microresonators make them well suited for close approach to a planar source mass, with most of their mass able to participate in the short-range interaction due to the thinness of the pad. Furthermore, the optical lever readout of their motion does not require additional, non-participating mass, as the laser beam reflects directly off of the pad surface~\cite{pratt2023nanoscale}. With mechanical quality factors exceeding one million due to dissipation dilution in the strained Si$_3$N$_4$ suspension, the devices reported in \cite{pratt2023nanoscale} possess room temperature thermal torque noise below $10^{-19}~\text{N}\,\text{m}/\sqrt{\text{Hz}}$, with the potential for further noise reduction through cryogenic cooling~\cite{geraci2008improved}. The devices may also be less susceptible to noise arising from electrostatic effects at narrow separations, as strain-induced stiffness minimizes the overall resonator motion~\cite{dong2023coupling}. Finally, lithographically defined geometry ensures precise dimensional control, improving fabrication tolerances while also enabling rapid production of multiple units to gather statistics on fabrication errors.

An example of the proposed experiment is depicted in Fig. \ref{fig:diagram}, where the gravitational coupling between a torsion microresonator and a heterogeneous source mass of comparable size is measured as a function of separation $s$. The source mass is continually rotated such that the gravitational torque on the pad oscillates at three times the rotation frequency (neglecting higher-order harmonics). Torque measurements at various separations can be used to distinguish a potential Yukawa interaction from a purely Newtonian gravitational signal, which would have a different $s$-dependence, as depicted in Fig. \ref{fig:ISL_projections}a. This procedure also helps discern contributions from separation-independent effects, such as wobble in the rotary system that produces vibrations at rotation frequency harmonics. To reduce electrostatic coupling between the source and test masses, an electrostatic shield is inserted between the two (omitted from Fig. \ref{fig:diagram}), such as a metallized Si$_3$N$_4$ membrane~\cite{geraci2008improved}. However, coupling between the resonator and the shield can still have effects such as altering the resonator stiffness or presenting a noise source at close separations~\cite{ke2023electrostatic,lee2020new,dong2023coupling}, potentially requiring the pad be metallized and grounded through partial metallization of the ribbon suspension~\cite{yu2012control}. Remaining contact potentials can be compensated by applying a bias voltage to the shield~\cite{tan2020improvement}. 

The experiment's sensitivity is fundamentally limited by thermal motion of the torsion resonator with a torque-equivalent power spectral density $S_\tau^\text{th}=8\pi k_\text{B}Tf_0 I_0 / Q_0$. While the moment of inertia $I_0=3\times 10^{-15}$ kg$\,$m$^2$ only depends on the pad dimensions, the resonance frequency $f_0$ and quality factor $Q_0$ depend also on the ribbon suspension geometry (see Appendix \ref{app:mechanicalModels}). The ribbon is assumed to have a $150~\upmu$m width, $40~$nm thickness, and 1.5 cm total length. Following Ref. \cite{pratt2023nanoscale}, these parameters are expected to yield $f_0=137$ Hz, $Q_0=6\times 10^7$, and $\sqrt{S_\tau^\text{th}}=3\times 10^{-20}~$N$\,$m/$\sqrt{\rm Hz}$ at $T=300~$K. Each torque measurement is polluted by thermal noise with standard deviation $\delta N=\sqrt{S_\tau^\text{th}/t_\text{meas}}$, depicted by dashed lines in Fig. \ref{fig:ISL_projections}a.

A nonzero Yukawa interaction would manifest as a deviation from the expected gravitational torque, so the minimum detectable interaction strength $\alpha_\text{min}$ is limited by uncertainty in both the total torque estimate and the Newtonian torque model. In particular, errors in calibration, fabrication, or alignment may lead to imperfect subtraction of the expected Newtonian contribution. To account for these effects, consider a model for the torque on the paddle with contributions from Newtonian gravity $\tau_\text{G}$ and a Yukawa interaction $\tau_\text{Y}$,
\begin{equation}
	\tau(s) = A \left(\tau_\text{G}(s,\boldsymbol{\beta}) + \tau_\text{Y}(s,\boldsymbol{\beta})\right) + B,
\end{equation}
where $\boldsymbol{\beta}$ is a vector whose components represent experimental parameters (including $\alpha$) that determine the gravitational signal and $A$ and $B$ are parameters encoding systematic error in the torque calibration. Multiple measurements at different locations $s$ distinguish the Yukawa torque from the Newtonian torque, inferring a best fit value for $\alpha$. However, each torque measurement will contain uncorrelated error due to thermal torque noise with standard deviation $\delta N$, such that a nonzero estimate of $\alpha$ may be inferred even in the absence of a Yukawa force. The expected variance of this estimate is used to define the minimum detectable interaction strength $\alpha_\text{min}$ to produce a statistically significant signal (see Appendix \ref{app:alphaMinAnalysis}). 

Generally, the Newtonian $\tau_\text{G}$ and Yukawa $\tau_\text{Y}$ signals depend on experimental parameters $\boldsymbol{\beta}$, such as those parametrizing the shape, size, or alignment of the source and test masses. In practice, the value of each parameter may not be known exactly and these uncertainties affect $\alpha_\text{min}$ (see Appendix \ref{app:alphaMinAnalysis} for more details). In Fig. \ref{fig:ISL_projections}b we account for uncertainty in the paddle widths $\delta w_x=\delta w_y=10~\upmu$m and uncertainty in the smallest separation $\delta s_0=5~\upmu$m. However, the limits presented are dominated by uncertainty in the torque signal calibration, where we have assumed an overall scale factor uncertainty of $\delta A=25\%$ and no prior knowledge of the offset $B$. Perfect knowledge of these parameters would reduce $\alpha_\text{min}\left(\lambda=30~\upmu\text{m}\right)$ for the 300 K, 300 K (tungsten), 4 K, and 100 mK curves in Fig. \ref{fig:ISL_projections}b by factors of 2.1, 3.1, 3.8, and 4.5, respectively.

\begin{figure}[b]
    \centering
    \includegraphics[width=1\columnwidth]{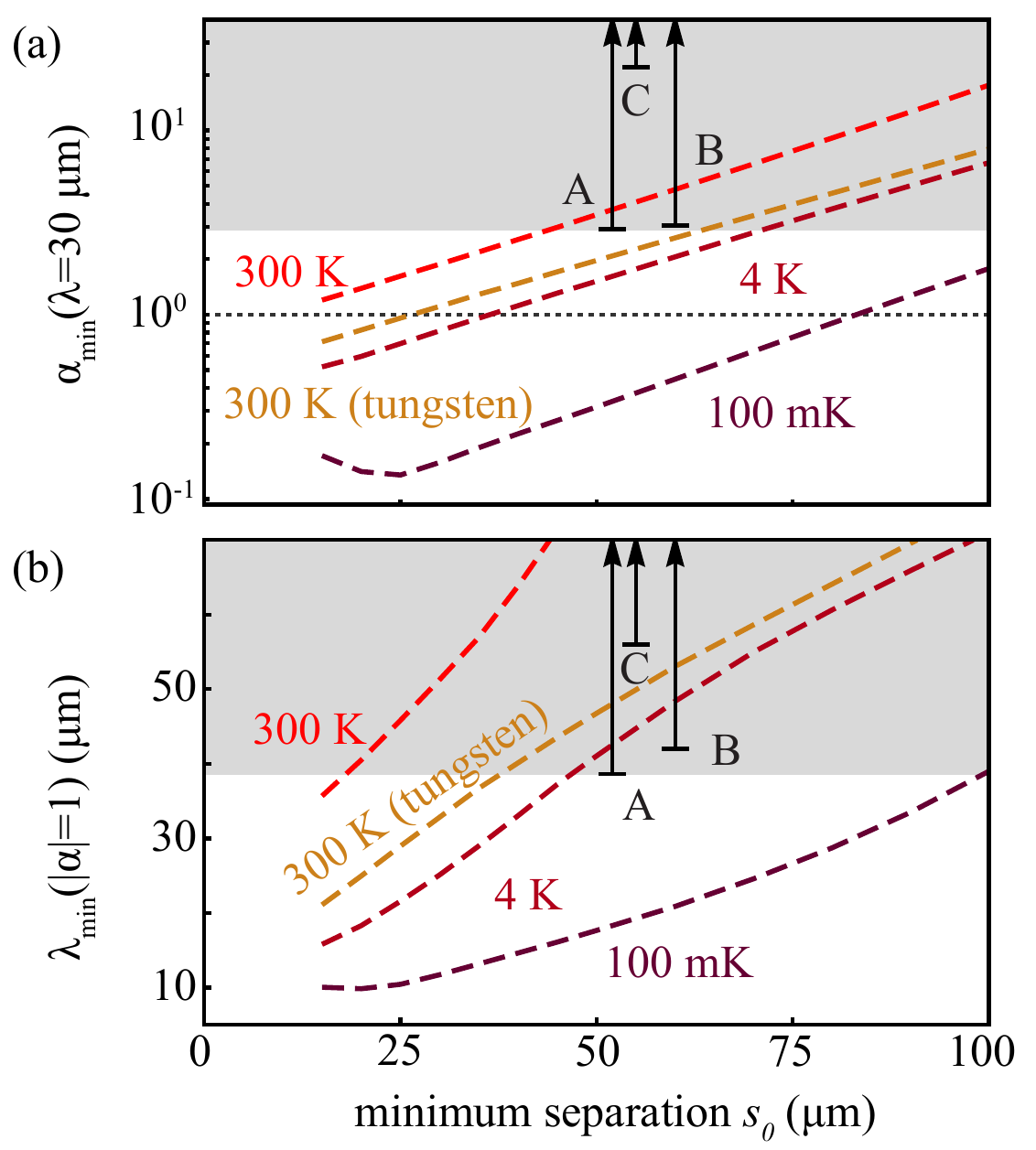}
    \caption{ \textbf{Projected performance versus minimum achievable surface separation.} Expected $2\sigma$ constraints are calculated as a function of the smallest separation distance $s_0$ achieved in the experiment. Measurements are assumed to be made on the interval $s\in\left[s_0,s_0+50~\upmu\text{m}\right]$ with all other parameters the same as in Figs. \ref{fig:diagram} and \ref{fig:ISL_projections}. (a) The minimum detectable interaction strength $\alpha_\text{min}$ at $\lambda=30~\upmu$m and (b) the minimum accessible interaction length $\lambda_\text{min}$ for a gravitational-strength Yukawa force ($\left|\alpha\right|=1$) are plotted as a function of $s_0$. The gray shaded region indicates the range of values for $\alpha_\text{min}(\lambda=30~\upmu\text{m})$ and $\lambda_\text{min}(\left|\alpha\right|=1)$ that have been excluded by other experiments, marked A~\cite{lee2020new}, B~\cite{Cook2013}, and C~\cite{Kapner2007}. }
	\label{fig:separation_plots}
\end{figure}

Figure \ref{fig:ISL_projections}b depicts the projected minimum detectable coupling strength $\alpha_\text{min}$ of a torsion resonator within the 1-300$~\upmu$m interaction range, including limitations from thermomechanical noise and systematic uncertainties. We find that a room temperature experiment consisting of 30, one-day-long measurements at separations in the range $s\in\left[25,75\right]~\upmu$m is capable of probing new parameter space over the interval $3~\upmu\text{m}\lesssim\lambda\lesssim36~\upmu$m (dashed red curve). To improve upon this result, additional measures can be taken such as cryogenic cooling to reduce thermal noise or fabricating the resonator from a denser material to amplify the Yukawa signal. These scenarios are also included in Fig. \ref{fig:ISL_projections}, where the thermal noise variance is simply rescaled by the temperature $\delta N^2\propto T$ for cryogenic cooling (4 K and 100 mK) and an additional room temperature projection is included assuming a tungsten test mass where all other device parameters are held constant, such that the tungsten-based device would have parameters $(I_0,f_0,Q_0)=(3\times 10^{-14}~\text{kg}\, \text{m}^2,\,48~\text{Hz},\,6\times 10^7)$.

\begin{figure*}[t]
    \centering
    \includegraphics[width=2\columnwidth,trim= 0in 0in 0in 0in]{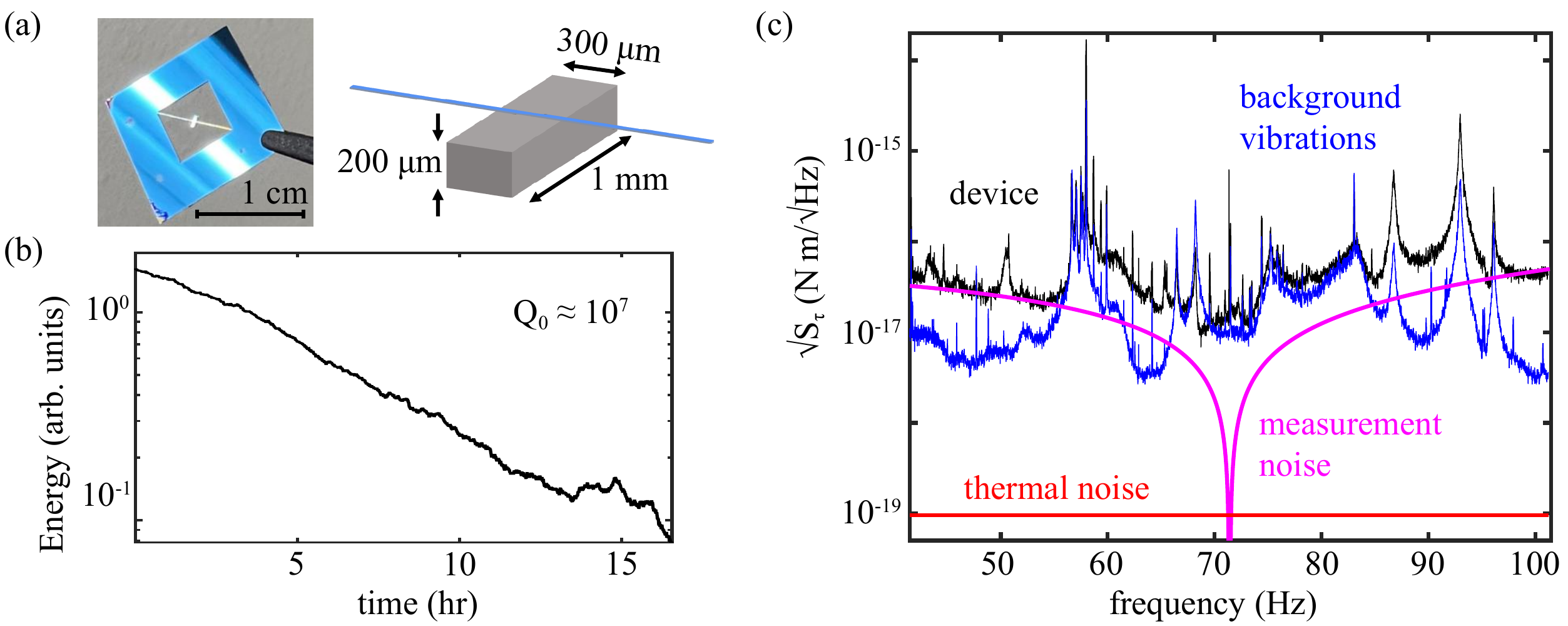}
    \caption{\textbf{Prototype device.} (a) Photograph and diagram of a prototype device. A Si$_3$N$_4$ ribbon with a width of 100$~\upmu$m and thickness of $80~$nm spans the $7~$mm gap of a window etched into a silicon chip. The ribbon is mass-loaded with a nominally 300$~\upmu$m $\times$ 1000$~\upmu$m $\times$ 200$~\upmu$m silicon pad. (b) Ringdown measurement of the torsion mode quality factor. (c) Measurement of the device's torque noise spectrum (black) reveals extraneous noise due to vibrations in the laboratory exceeding the expected thermal noise (red) by two orders of magnitude. The measurement contains additional readout noise with an approximately white angular displacement spectrum of ${\sim 18}$ nrad/$\sqrt{\rm Hz}$, modeled by the magenta curve. A concurrent measurement using a commercial seismometer is included (blue), converted to an inferred torque noise via $\sqrt{S_\tau}=\left(m_0 w_z/2\right) \sqrt{S_a}$. } 
	\label{fig:prototypeDevice}
\end{figure*}

The minimum separation distance in the experiment may be limited by electrostatic coupling between the resonator and the electrostatic shield. Metallization and voltage compensation can reduce this effect~\cite{tan2020improvement}, however, nonuniform potentials on the conducting surfaces will remain and present a statistical, separation-distance-dependent noise source in the form of seismic patch field coupling~\cite{dong2023coupling,lee2020fourier}. Polycrystallinity, surface contamination, or variation in chemical composition produce a nonuniform surface potential, commonly referred to as ``patch potentials"~\cite{speake2003forces}. Random translational or rotational motion of the oscillator along any axis causes it to sample the spatially random and anharmonic potential established by the patch fields, resulting in random forces and torques. Due to the short range of the patch fields whose length scale is assumed to be smaller than the separation distance $s$, this effect is more pronounced at shorter distances and is expected to scale as $s^{-4}$~\cite{behunin2012modeling,turchette2000heating}. The effect of this seismic patch field coupling can be reduced by decreasing the oscillator's motion through a combination of vibration isolation and simultaneous feedback cooling~\cite{poggio2007feedback} of the torsional and flexural modes. 

The exact limitations on our experiment posed by electrostatic effects are unknown but will likely manifest as a lower limit on the surface separation. Unsure of this limit in the proposed system, in Fig. \ref{fig:separation_plots} we explore the effect of varying the smallest separation distance $s_0$ on $\alpha_\text{min}$ and the minimum interaction length $\lambda_\text{min}$ for which the experiment is sensitive to gravitational-strength Yukawa couplings ($\left|\alpha\right|=1$). For a room temperature silicon device (red), Fig. \ref{fig:separation_plots}a shows that for $\lambda=30~\upmu$m a minimum separation $s_0\lesssim 45~\upmu$m is needed to surpass prior constraints and $s_0\lesssim 10~\upmu$m is needed to achieve $\alpha_\text{min}\leq 1$. Figure \ref{fig:separation_plots}b shows that new constraints on $\lambda_\text{min}$ would require $s_0\lesssim 18~\upmu$m, while a cryogenic experiment at 100 mK (purple) could potentially achieve $\lambda_\text{min}\approx 10~\upmu$m with $s_0\approx 25~\upmu$m.

As a first step toward experiment, we have fabricated and characterized a prototype microresonator with dimensions comparable to the proposed devices. For the design dimensions in Fig. \ref{fig:prototypeDevice}a, the predicted frequency, quality factor, and thermal torque noise are 80 Hz, $1.3\times10^7$, and $10^{-19}~$N$\,$m/$\sqrt{\rm Hz}$, respectively (see Appendix \ref{app:mechanicalModels}). Calibrated readout of the resonator's angular displacement was performed using an optical lever~\cite{pratt2023nanoscale} with a $1550~$nm laser beam, and Fig. \ref{fig:prototypeDevice}b shows a ringdown measurement that confirms the design quality factor $Q_0\approx 10^7$. The displacement spectrum $S_\theta$ reveals a lower resonance frequency $f_0=71.5$ Hz than predicted, possibly due to overestimation of the thin-film stress, for example. The apparent torque spectrum (Fig. \ref{fig:prototypeDevice}c) can be inferred as $S_\tau = \left|\chi\right|^{-2} S_\theta$, where the inverse mechanical susceptibility is $\chi^{-1}\equiv \left(2\pi\right)^2 I_0 \left({f_0}^2-f^2-i f f_0 / Q_0\right)$. As shown in Fig. \ref{fig:prototypeDevice}c, we infer $\sqrt{S_\tau}\approx10^{-17}~$N$\,$m/$\sqrt{\rm Hz}$.

The torque noise of our prototype device (Fig. \ref{fig:prototypeDevice}c) is two orders of magnitude above the predicted thermal torque noise and points to a key challenge: mitigating acceleration noise that couples to the torsion mode due to asymmetric mass loading. Specifically, the excess noise corresponds to a horizontal acceleration background of 70 n$g$/$\sqrt{\rm Hz}$ near 71.5 Hz, which was confirmed with an independent accelerometer measurement (blue trace in Fig. \ref{fig:prototypeDevice}c). We note that the current device is particularly susceptible to horizontal accelerations given its relatively large ($w_z=200~\upmu$m) thickness, since $\sqrt{S_\tau}\propto {w_z}^2 \sqrt{S_a}$. By reducing the device thickness to 50$~\upmu$m, per the proposed design, this vibration-induced torque noise would be reduced by a factor of 16.  To completely remove the coupling of horizontal accelerations into the torsion mode, the device may be fabricated with the center of mass aligned with the torsion axis, which can be achieved by depositing additional material on the topside of the pad, for example. 

In summary, we have proposed microscale torsion resonators~\cite{pratt2023nanoscale} as a new platform for short-range gravity experiments and modeled the expected performance given limitations from thermomechanical noise and systematic uncertainties. We find that a room temperature device has sufficient sensitivity to detect new Yukawa interactions below ${\sim 36}~\upmu$m, assuming surface separations down to 25$~\upmu$m and a 30-day-long measurement campaign. As a first step, we have fabricated a prototype device exhibiting a large mechanical quality factor of $10^7$ and thermal torque of $10^{-19}~$N$\,$m/$\sqrt{\rm Hz}$. The key next steps are addressing acceleration noise and potential electrostatic interactions such that measurements can be performed near the device thermal limit. In addition to new tests of Yukawa interactions, this would enable the first measurements of the Newtonian gravitational coupling between sub-milligram objects~\cite{westphal2021measurement}. Looking forward, we note that microscale torsion resonators are a promising platform for exploring new physics beyond ISL deviations, such as spin-dependent interactions~\cite{terrano2015short} or quantum gravity~\cite{carney2021using}. 

We thank Patrick Egan and Charles Clark for feedback on the manuscript. This work is supported by the National Science Foundation Grants No. 2239735 and No. 2134830 and Arizona State University.

\appendix

\section{Analysis for minimum detectable coupling strength} \label{app:alphaMinAnalysis}
The proposed experiment will use the gravitational torque exerted on the torsion paddle by the source mass to infer a value $\hat{\alpha}$ of the Yukawa coupling strength $\alpha$. The inferred coupling strength $\hat{\alpha}$ contains error from both stochastic noise and systematic uncertainties. The non-zero variance $\delta \hat{\alpha}$ of this estimate affects the minimum detectable coupling strength $\alpha_\text{min}$. 

In this section, we provide a treatment for calculating $\alpha_\text{min}$ based on estimated limitations from thermal noise and systematic error due to uncertainty in experimental parameters. We also provide a brief explanation of the numerical simulations used to model the expected Newtonian and Yukawa torque signals.

\subsection{Measurement and parameter uncertainty}
At a given surface separation distance $s_i$, the source mass exerts a torque on the paddle containing both Newtonian $\tau_G$ and Yukawa $\tau_Y$ contributions
\begin{equation}
	\tau_{\text{grav},i}(\boldsymbol{\beta}) =\tau_G(\boldsymbol{\beta},\Delta s_i) + \tau_Y(\boldsymbol{\beta},\Delta s_i).
\end{equation}
where $\boldsymbol{\beta}$ is a vector whose components represent the experimental parameters that determine this torque signal.\footnote{While the interaction strength $\alpha$ is included as a component of $\boldsymbol{\beta}$, the interaction length $\lambda$ is not. The value of $\lambda$ is assumed, and the calculations for $\alpha_\text{min}$ presented here must be repeated separately for each value of $\lambda$ within the range of interest.} The separation distances have been parametrized as $s_i = s_0 + \Delta s_i$, such that the increments $\Delta s_i$ are treated as the independent variables and the smallest separation $s_0$ is included as a component in $\boldsymbol{\beta}$. 

Experimentally, $\tau_{\text{grav},i}$ is not measured directly, but rather it is inferred from some experimentally accessible quantity, such as the photocurrent output from a photodetector. With this in mind, we define the function
\begin{equation}
	v_i(\boldsymbol{\eta}) = a\, \tau_{\text{grav},i}(\boldsymbol{\beta}) + b,
\end{equation}
to model the raw data (ignoring stochastic noise in the measurement), which is related to the signal $\tau_{\text{grav},i}$ by a calibration factor $a$. Another parameter $b$ accounts for an overall offset in the measurement. A more general parameter vector $\boldsymbol{\eta}\equiv\left(\beta_0,\beta_1,...,a,b\right)$ includes these components $(a,b)$ in addition to the parameters in $\boldsymbol{\beta}$.

The actual measured data
\begin{equation} \label{eq:app:measuredVoltage}
	v_{\text{meas},i}(\boldsymbol{\eta},\bar{v}_{N,i}) = v_i(\boldsymbol{\eta}) + \bar{v}_{N,i}
\end{equation}
contain stochastic noise terms $\bar{v}_{N,i}$, which we treat as independent, mean-zero random variables\footnote{Here, independent random variables are given overbars (e.g. $\bar{x}$) while fit parameters that will be inferred from the data are given hats (e.g. $\hat{x}$).} with standard deviation $\delta v_{N,i}$.

In an experiment, the parameters $\boldsymbol{\eta}$ may not be precisely known and must be inferred by fitting a model $v_i(\hat{\boldsymbol{\eta}})$ with fit parameters $\hat{\boldsymbol{\eta}}$ to the experimental data $v_{\text{meas},i}(\boldsymbol{\eta},\bar{v}_{N,i})$. The fit parameters can be additionally constrained by independent measurement or estimated from the design specification. Given measurement error or finite fabrication tolerance, the measured values $\bar{{\eta}}_\nu$ are treated as random variables, each with mean $\eta_\nu$ and estimated standard deviation ${\delta \eta_\nu}$.  

Following the treatment in Ref. \cite{lee2020new}, the data from an experiment with multiple measurements can be fitted to estimate the value of $\boldsymbol{\eta}$ by minimizing the quantity
\begin{equation} \label{eq:chi2}
	\chi^2 = \sum_i \frac{\left( v_{\text{meas},i}(\boldsymbol{\eta},\bar{v}_{N,i}) - v_i(\hat{\boldsymbol{\eta}})\right)^2}{{\delta v_{N,i}}^2} + \sum_\nu \frac{\left(\bar{\eta}_\nu - \hat{\eta}_\nu \right)^2}{{\delta\eta_{\nu}}^2}
\end{equation}
with respect to $\boldsymbol{\hat{\eta}}$. Here, Latin indices refer to torque measurements and Greek indices refer to individual parameters and their independent measurements or estimates. 

Our analysis in the main text accounts for uncertainty in several experimental parameters: the Yukawa interaction strength ($\alpha$), the torsion resonator widths ($w_x$,$w_y$), and the smallest source-test mass separation distance ($s_0$), such that $\boldsymbol{\hat{\beta}}=\left(\hat{\alpha},\hat{w}_x,\hat{w}_y,\hat{s}_0\right)$ and $\boldsymbol{\hat{\eta}}=\left(\hat{\alpha},\hat{w}_x,\hat{w}_y,\hat{s}_0,\hat{a},\hat{b}\right)$. While the vector $\boldsymbol{\beta}$ will more generally contain additional parameters, such as the material densities or geometric parameters of the source mass, for simplicity we assume they are known with certainty. Note that this choice of parameters means that we include uncertainty in $s_0$, but we treat the increments $\Delta s_i$ as independent variables with negligible error.

\subsection{Linearized solution}
For a prospective experiment, the variance ${\delta \hat{\eta}_\nu}^2$ of the fit parameters can be estimated in terms of the expected noise level $\delta v_{N,i}$ and parameter uncertainties $\delta \eta_\nu$. For convenience, we start by substituting the parameters
\begin{equation}
	\begin{aligned}
		A=1,& \,\,\,\,\, &B=b/a&, \\
		\hat{A}=\hat{a}/a,& &\hat{B}=\hat{b}/a&, \\
		\bar{A}=\bar{a}/a,& &\bar{B}=\bar{b}/a&,
	\end{aligned}
\end{equation}
and redefining the parameter vector accordingly: $\boldsymbol{\eta}\rightarrow \left(\beta_0,\beta_1,...,A,B\right)$. By defining a new function
\begin{equation}
 	\tau_i(\boldsymbol{\eta})= A\, \tau_{\text{grav},i}(\boldsymbol{\beta})+B,
\end{equation}
Eq. \ref{eq:chi2} transforms to 
\begin{equation}
	\chi^2 = \sum_i  \frac{\left( \tau_{i}(\boldsymbol{\eta}) + \bar{N}_i - {\tau}_i(\boldsymbol{\hat{\eta}})\right)^2}{{\delta N_i}^{2}} + \sum_\nu \frac{\left(\bar{\eta}_\nu - \hat{\eta}_\nu \right)^2}{{\delta \eta_\nu}^2},
\end{equation}
where $\bar{N}_i=\bar{v}_{N,i}/a$ is the torque-equivalent noise. In the main text we assume this noise is due to thermal torque noise in the resonator with variance ${\delta {N}_i}^2=S_\tau^\text{th}/t_\text{meas}$. 

An analytical solution for the fit parameters $\hat{\eta}_\nu$ that minimize $\chi^2$ can be obtained by linearizing $\tau_i(\boldsymbol{\hat{\eta}})$ with respect to its parameters
\begin{equation}
	\begin{aligned}
		\tau_i(\hat{\boldsymbol{\eta}}) &\approx \tau_i({\boldsymbol{\eta}}) + \sum_\nu \left( \hat{{\eta}}_\nu - {\eta_\nu}\right) \Psi_{i\nu} \\
		\Psi_{i\nu} &\equiv \left( \left. \frac{\partial \tau_i(\hat{\boldsymbol{\eta}})}{\partial \hat{\eta}_\nu}  \right|_{\hat{\boldsymbol{\eta}} = \boldsymbol{\eta}} \right)
	\end{aligned} 
\end{equation}
Under this approximation, it can be shown that the solution for each component of $\hat{\boldsymbol{\eta}}$ will have an expected variance 
\begin{equation}
	{\delta {\hat{\eta}_\nu}}^2 = {\delta \eta_\nu}^{2} D_{\nu\nu}
\end{equation}
where
\begin{equation}
	\begin{aligned}
		\overleftrightarrow{\boldsymbol{D}}\equiv & \overleftrightarrow{\boldsymbol{C}}^{-1} \\
		C_{\mu\nu}\equiv	&	\delta_{\mu\nu} +\sum_{i} {\delta {N}_i}^{-2} {\delta \eta_\mu}\delta \eta_\nu\Psi_{i\mu} \Psi_{i\nu} \\
	\end{aligned}
\end{equation}
We define the minimum detectable coupling strength $\alpha_\text{min}$ in terms of the expected variance of the estimator $\hat{\alpha}$ under the assumption that $\alpha=0$ and that $\hat{\alpha}$ is not constrained by prior measurement ($\delta \alpha=\infty$). If the 0th parameter is $\alpha$, i.e. $\eta_0 = \alpha$, then for 2$\sigma$ confidence
\begin{equation}
	\alpha_\text{min} \equiv 2 \,\delta {\hat{\alpha}} = \left.2\delta \alpha \sqrt{D_{00}}\right|_{\delta \alpha=\infty}
\end{equation}

\subsection{Torque simulations}
We performed numerical simulations to estimate the amplitudes of the torques $\tau_\text{G}({\boldsymbol{\eta}},\Delta s)$ and $\tau_\text{Y}({\boldsymbol{\eta}},\Delta s)$ exerted on the test mass by the source mass. For various interaction lengths in the range $\lambda\in\left[1~\upmu \text{m},1 \text{ cm}\right]$, the Newtonian and Yukawa potentials were calculated over a rectangular grid in the region $x\in$ [-550,550]$~\upmu$m, $y\in$ [-550,550]$~\upmu$m, $z\in$ [25,250]$~\upmu$m, using stratified Monte Carlo sampling of the source mass distribution. A numerical gradient operation was then performed to extract the vector Newtonian and Yukawa field components. 

The torque on the torsion resonator was computed for a given pad geometry as a weighted (accounting for the local lever arm) numerical integration over the Newtonian and Yukawa fields. These torque calculations were repeated at various surface separations to infer the torque's dependence on separation distance $s$. Parameters $\left(w_x,w_y\right)$ were also independently swept at each separation distance in order to calculate $\Psi_{i\nu}$. In order to estimate $\alpha_\text{min}$ for any arbitrary set of chosen separation distances $s_i$, polynomial fits were performed to approximate the sampled $\Psi_{i\nu}$ as smooth functions of separation distance.

\section{Models for the resonator mechanical properties} \label{app:mechanicalModels}
The thermal torque noise generally depends on the resonator's frequency $f_0$, moment of inertia $I_0$, and quality factor $Q_0$. Here we present the models used for each of these parameters. The frequency can be modeled as
\begin{equation}
    f_0 = \frac{1}{2\pi}\sqrt{\frac{k_E + k_\sigma + k_g}{I_0}}
\end{equation}
where $I_0= \rho_\text{pad} w_x w_y w_z \left({w_y}^2/12+{w_z}^2/3 \right)$ and $k_E$ ($k_\sigma , k_g$) is the elastic (tensile, gravitational) torsional stiffness~\cite{pratt2023nanoscale}. The pad's density is $\rho_\text{pad}$. We model the elastic and tensile stiffness as~\cite{pratt2023intersection}
\begin{equation}
    \begin{aligned}
	   k_E &= \frac{2E_\text{rib} {h_\text{rib}}^3 w_\text{rib}}{3 L_\text{rib}}\\
	   k_\sigma &=\frac{\sigma_\text{rib} {h_\text{rib}} {w_\text{rib}}^3}{3 L_\text{rib}}. 
    \end{aligned}
\end{equation}
where $w_\text{rib}$, $L_\text{rib}$, and $h_\text{rib}$ are respectively the ribbon's width, length, and thickness. Following Ref.~\cite{pratt2023nanoscale}, we assume the ribbon's stress to be $\sigma_\text{rib}=0.85$ GPa and elastic modulus to be $E_\text{rib}=250$ GPa. The gravitational stiffness comes from the restoring torque exerted on the torsion pad by Earth's gravity $g$. When the device is oriented such that the pad hangs below the ribbon, the gravitational stiffness is 
\begin{equation}
	k_g = 	\frac{1}{2} g \rho_\text{pad} w_x w_y {w_z}^2.
\end{equation}
Mechanical dissipation in the resonator inherently depends on the material's intrinsic quality factor, which for Si$_3$N$_4$ can be modeled as~\cite{villanueva2014evidence}
\begin{equation}
        Q_\text{i} = 60 \left(\frac{h_\text{rib}}{1~\text{nm}}\right) 
\end{equation}
where surface loss plays a larger role in thinner films. However, due to dissipation dilution, the quality factor of the resonator $Q_0$ is enhanced as~\cite{pratt2023nanoscale}
\begin{equation}
	Q_0 = Q_\text{i} \left(1 + \frac{k_\sigma + k_g}{k_E}\right).
\end{equation}

\bibliography{references.bib}
\end{document}